\documentclass[preprint1]{aastex}
\usepackage{graphicx}
\usepackage{epsfig}
\usepackage{amssymb,amsmath}
\usepackage{rotating}

\slugcomment{To appear in ApJ}
\shorttitle{A peculiar cataclysmic variable}
\shortauthors{Rebassa-Mansergas et al.}

\newcommand{\Mwd}{\mbox{$M_\mathrm{wd}$}}

\newcommand{\Msun}{\mbox{$\mathrm{M}_{\odot}$}}

\newcommand{\Porb}{\mbox{$P_{\mathrm{orb}}$}}
\newcommand{\Line}[3]{\Ion{#1}{#2}\,$\lambda$\,#3}
\newcommand{\Lines}[3]{\Ion{#1}{#2}\,$\lambda\lambda$\,#3}
\newcommand{\Ion}[2]{#1{\,\scriptsize #2}}

\begin{document}

\title{SDSS\,J001153.08-064739.2, a cataclysmic variable with an evolved donor in the period gap{\LARGE{$^{\star}$}}}

\author{A. Rebassa-Mansergas\altaffilmark{1},
S.G. Parsons\altaffilmark{2},
C.M. Copperwheat\altaffilmark{3},
S. Justham\altaffilmark{4},
B.T. G\"ansicke\altaffilmark{5},\\
M.R. Schreiber\altaffilmark{2,6},
T.R. Marsh\altaffilmark{5}, 
V.S. Dhillon\altaffilmark{7}
}

\altaffiltext{}{$\star$ Based in part on observations made with NTT/Ultracam under program 087.D-0557, and Magellan Clay/MIKE.}
\altaffiltext{1}{Kavli Institute for Astronomy and Astrophysics, Peking University, Beijing 100871, PR China.} 
\altaffiltext{2}{Departamento de F\'{\i}sica y Astronom\'{\i}a, Universidad de Valpara\'{\i}so, Valpara\'{\i}so, Chile.}
\altaffiltext{3}{Astrophysics Research Institute, Liverpool John Moores University, IC2, Liverpool Science Park, 146 Brownlow Hill, Liverpool L3 5RF, UK.}
\altaffiltext{4}{National Astronomical Observatories, Chinese Academy of Sciences, 100012 Beijing, PR China.}
\altaffiltext{5}{Department of Physics, University of Warwick, Coventry CV4 7AL, UK.}
\altaffiltext{6}{Millennium Nucleus "Proto-planetary Disks in ALMA Early Science", Universidad de Valpara\'\i so, Avenida Gran Breta\~na 1111, Valpara\'\i so, Chile.}
\altaffiltext{7}{Department of Physics and Astronomy, University of Sheffield, Sheffield, S3 7RH.}

\begin{abstract}

Secondary stars in  cataclysmic variables (CVs) follow  a well defined
period-density  relation.   Thus, canonical  donor  stars  in CVs  are
generally low-mass  stars of  spectral type  M.  However,  several CVs
have been  observed containing secondary  stars which are too  hot for
their inferred masses.  This particular configuration can be explained
if  the donor  stars in  these systems  underwent significant  nuclear
evolution  before they  reached  contact.  In  this  paper we  present
SDSS\,J001153.08-064739.2 as  an additional example belonging  to this
peculiar type  of CV and  discuss in detail its  evolutionary history.
We perform spectroscopic and photometric  observations and make use of
available Catalina  Real-Time Transient  Survey photometry  to measure
the  orbital  period of  SDSS\,J001153.08-064739.2  as  2.4 hours  and
estimate   the  white   dwarf   ($\Mwd>0.65\Msun$)   and  donor   star
($0.21\Msun<M_\mathrm{don}<0.45\Msun$)  masses, the  mass ratio  ($q =
0.32 \pm 0.08$), the orbital  inclination ($47^\circ < i < 70^\circ$),
derive  an  accurate  orbital  ephemeris ($T_{0}  =  2453383.578(1)  +
\mathrm{E}\times0.10028081(8)$),  and  report   the  detection  of  an
outburst.  We show  that SDSS\,J001153.08-064739.2 is one  of the most
extreme cases in which the donor star is clearly too hot for its mass.
SDSS\,J001153.08-064739.2  is   therefore  not  only  a   peculiar  CV
containing an evolved  donor star but also an accreting  CV within the
period   gap.   Intriguingly,   approximately   half   of  the   total
currently-observed sample  of these  peculiar CVs  are located  in the
period gap with nearly the same orbital period.

\end{abstract}


\section{Introduction}
\label{intro}

A cataclysmic variable (CV) is a close compact binary star formed by a
white dwarf  primary and a low-mass  near-main-sequence secondary star
(see \citealt{kniggeetal11-1}  for a recent review).   In non-magnetic
CVs the  companion transfers material  to the white dwarf  primary via
Roche-lobe overflow and  an accretion disk is formed  around the white
dwarf.

The  orbital periods  (\Porb)  of  CVs decrease  as  a consequence  of
orbital  angular momentum  loss  (AML). The  two  mechanisms that  are
thought  to drive  AML are  magnetic wind  braking of  the donor  star
\citep{verbunt+zwaan81-1}       and      gravitational       radiation
\citep{paczynski81-1}, the former several orders of magnitude stronger
than the latter.   Once the secondary becomes  fully convective (\Porb
$\sim$    3   hours)    magnetic    braking    is   greatly    reduced
\citep{rappaportetal83-1,                           schreiberetal10-1,
  rebassa-mansergasetal13-1} causing the secondary  to detach from its
Roche-lobe in the  2--3 hour orbital period range known  as the period
gap    (see    the    orbital     period    distribution    by    e.g.
\citealt{gaensickeetal09-1}).   At this  stage  the evolution  becomes
much slower, as  the AML is mainly driven  by gravitational radiation.
At  \Porb $\sim$  2 hours  the  shrinking Roche-lobe  comes back  into
contact with the secondary star and mass transfer is resumed at a much
lower rate.  Once the secondary  star becomes degenerate (\Porb $\sim$
80  minutes), the  overall  mass-radius relation  for  the donor  star
changes such  that it becomes  larger upon mass loss.   Therefore, the
orbital period  increases with  further mass  transfer and  the binary
orbit widens \citep{rappaportetal82-1, littlefairetal06-2}.

The secondary  stars of CVs  with orbital periods below  $\sim6$ hours
are  apparently un-evolved  M-dwarf main  sequence stars,  as expected
from  the well  known orbital  period-density relation  for Roche-lobe
filling stars \citep{kniggeetal11-1}.  However,  several CVs have been
recently identified  harbouring secondary stars  that are too  hot for
their      masses:      V\,485      (\Porb     =      59      minutes,
\citealt{augusteijnetal96-1}), EI\,PSc ( also known as RXJ\,2329+0628,
\Porb  =  64  minutes,  \citealt{thorstensenetal02-1}),  QZ\,Serpentis
(\Porb      =       2      hours,      \citealt{thorstensenetal02-2}),
SDSS\,J013701.06-091234.9       (\Porb       =       84       minutes,
\citealt{imadaetal06-1}),   SDSS\,J170213.26+322954.1  (\Porb   =  2.4
hours,  \citealt{littlefairetal06-1}), CSS\,J134052.0+151341  (\Porb =
2.45  hours,   \citealt{thorstensen13-1}),  GALEX\,J194419.33+491257.0
(\Porb =  76 minutes, \citealt{kato+osaki14-1}). A  few examples exist
of longer  orbital period  CVs that contain  secondary stars  that are
also  too  hot for  their  masses  (AE\,Aquarii,  \Porb =  9.9  hours,
\citealt{schenkeretal02-1};   HS\,0218+3229,   \Porb  =   7.1   hours,
\citealt{rodriguez-giletal09-1}).   It  has  been suggested  that  the
current peculiar  configuration of these  systems can be  explained if
the  secondary  stars  underwent  considerable  nuclear  evolution  --
i.e. they  were nearing the  end of the  main sequence --  before they
began to transfer mass onto the white dwarfs \citep{beuermannetal98-1,
  baraffe+kolb00-1,  podsiadlowskietal03-1}.  In  this scenario  these
more evolved donor  stars were initially more massive  than typical CV
donors and likely underwent a  phase of intense thermal-timescale mass
transfer. Under  these circumstances,  the accreted material  from the
evolved donors should  reflect in a different  composition compared to
canonical  CVs, i.e.   due  to CNO-processes  we  expect depletion  of
carbon  versus   nitrogen.   Indeed,   unusually  high   \Ion{N}{V}  /
\Ion{C}{IV} line  flux ratios  have been observationally  confirmed in
ultraviolet  spectra  of \citep{gaensickeetal03-1}  and  QZ\,Serpentis
\citep{gaensicke04-1}.  However,  it has  to be  stressed that  such a
phase  of mass  transfer  is not  the underlying  cause  of the  later
unusual appearance of the donor stars, as many canonical CVs have also
potentially passed through a phase  of thermal timescale mass transfer
\citep{schenkeretal02-1}.

\begin{figure}[th]
\centering
\includegraphics[width=8cm, angle=-90]{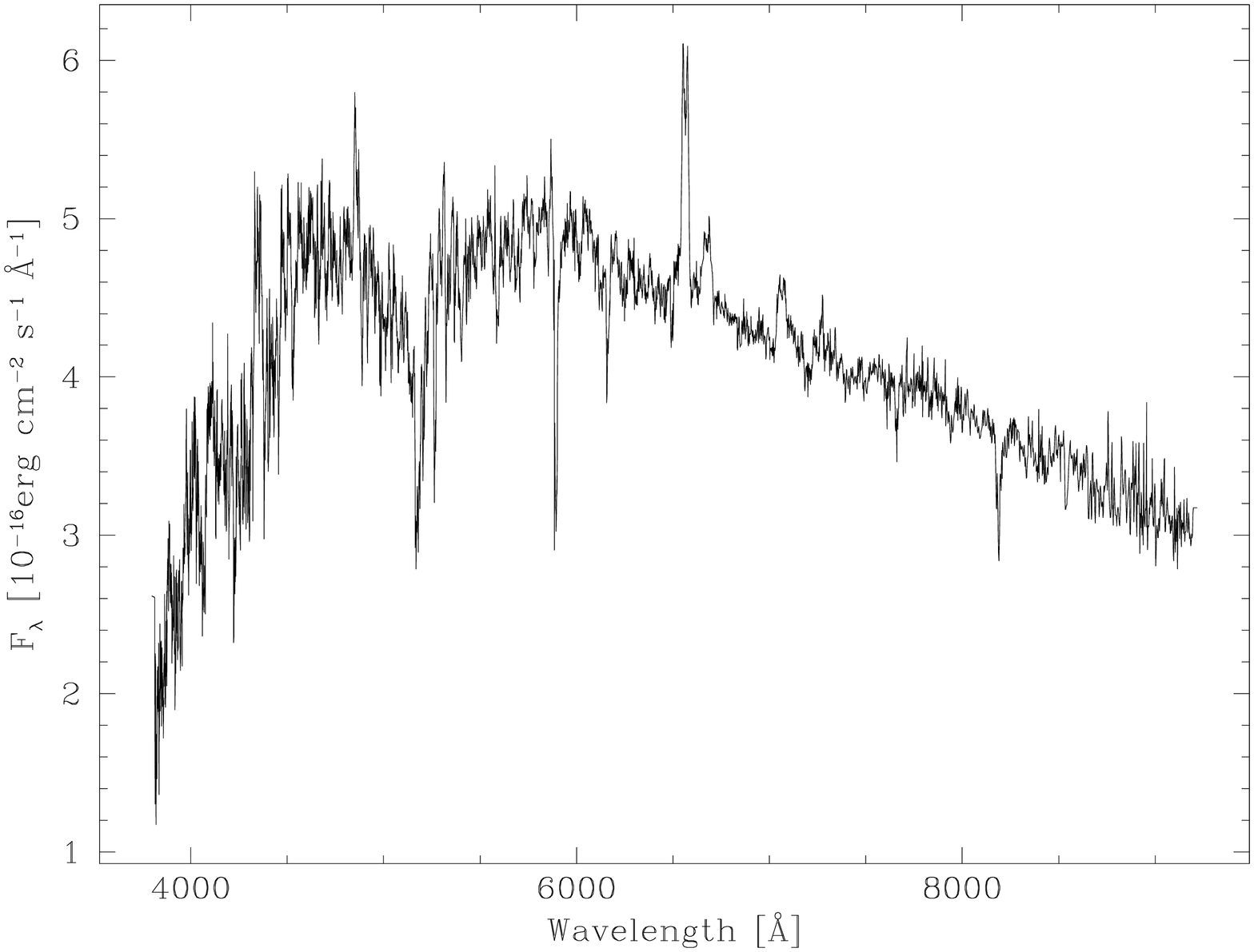}
\caption{SDSS  spectrum  of  SDSS\,J0011-0647 revealing  the  spectral
  features of the  donor star and double-peaked  Balmer emission lines
  characteristic  of  an  accretion  disk.   No  obvious  white  dwarf
  features can be seen.}
\label{f-spec}
\end{figure}

In the  course of  our search for  white dwarf-main  sequence binaries
\citep{rebassa-mansergasetal10-1,           rebassa-mansergasetal12-1,
  rebassa-mansergasetal13-2} in  the Sloan  Digital Sky  Survey (SDSS;
\citealt{abazajianetal09-1,     aiharaetal11-1})     we     discovered
SDSS\,J001153.08-064739.2  (hereafter  SDSS\,J0011-0647) and  we  here
present this object as a new system belonging to this peculiar type of
CV. We show that SDSS\,J0011-0647 is  one of the most extreme cases in
which  the secondary  star is  far  too hot  for its  mass.  The  SDSS
spectrum of  SDSS\,J0011-0647 is  shown in  Figure\,\ref{f-spec}.  The
overall  shape  of the  spectrum  is  blackbody-like, however  reveals
broad,  double  peaked Balmer  emission  lines,  characteristic of  an
accretion disk, and a very  strong \Ion{Mg}{I} absorption feature near
5150\AA.  This latter feature is  the main absorption complex in mid-K
stars, which we  attribute to the donor star  of SDSS\,J0011-0647.  No
obvious  white dwarf  features however  can be  identified.  The  SDSS
$ugriz$ magnitudes of SDSS\,J0011-0647 are 19.0, 17.8, 17.1, 16.9, and
16.8 respectively.

\section{Observations}
\label{obs}

In this section we briefly describe the telescope/instrumentation used
for the  follow-up observations  of SDSS\,J0011-0647, and  outline the
reduction of the data.

\subsection{Spectroscopy}

Time-resolved  high-resolution  spectroscopy of  SDSS\,J0011-0647  was
obtained  along  the 25th  of  June  2011 and  the  22nd  and 23rd  of
September  2012  with  the  Magellan Inamori  Kyocera  Echelle  (MIKE)
spectrograph on the  6.5m Clay Telescope at  Las Campanas Observatory.
The CCD  of MIKE's  red arm  has a  pixel scale  of 0.13\arcsec/pixel,
therefore  we  binned  the  detector  by a  factor  of  three  in  the
dispersion direction  and a  factor of two  in the  spatial direction,
thus reducing the readout time  and readout noise. This, together with
a  1\arcsec\,slit, provided  access to  the 5200-9000\AA\,  wavelength
range at a resolving power of 22,000 (i.e. 0.3\AA\, at the location of
the  H$\alpha$  line).   A  total  of 20  spectra  were  obtained  for
SDSS\,J0011-0647,  with  exposure  times  of 600  seconds.   The  data
analysis (reduction  and wavelength calibration) was  carried out with
the $IRAF$ software using the standard $IRAF$ package $IMRED:ECHELLE$.

\subsection{Photometry}
\label{s-phot}

SDSS\,J0011-0647  was observed  with  the  high-speed camera  ULTRACAM
\citep{dhillonetal07-1}  mounted as  a visitor  instrument on  the ESO
3.5m New Technology Telescope (NTT) at La Silla, Chile on the 24th and
25th  of May  2011. ULTRACAM  consists  of three  frame transfer  CCDs
allowing  one to  obtain  simultaneous photometry  in three  different
filters. For  our observations of  SDSS\,J0011-0647 we used  the $u'$,
$g'$ and  $r'$ filters.  On both  nights the seeing was  poor ($>$2'')
and the  moon virtually full,  hence we used relatively  long exposure
times  of 10  seconds (the  dead-time between  the 10  second ULTRACAM
exposures  was  only  25\,ms).  We  obtained  just  over  an  hour  of
photometry  on the  first night  and almost  two hours  on the  second
night.

All the  data were bias  subtracted, flat-fielded and  extracted using
the  ULTRACAM pipeline  \citep{dhillonetal07-1}. The  source flux  was
determined with aperture photometry using a variable aperture, whereby
its radius  was scaled according  to the  full width at  half maximum.
Variations in  observing conditions were accounted  for by determining
the flux relative to several comparison stars in the field of view.

\begin{figure}[th]
\centering
\includegraphics[width=7cm, angle=-90]{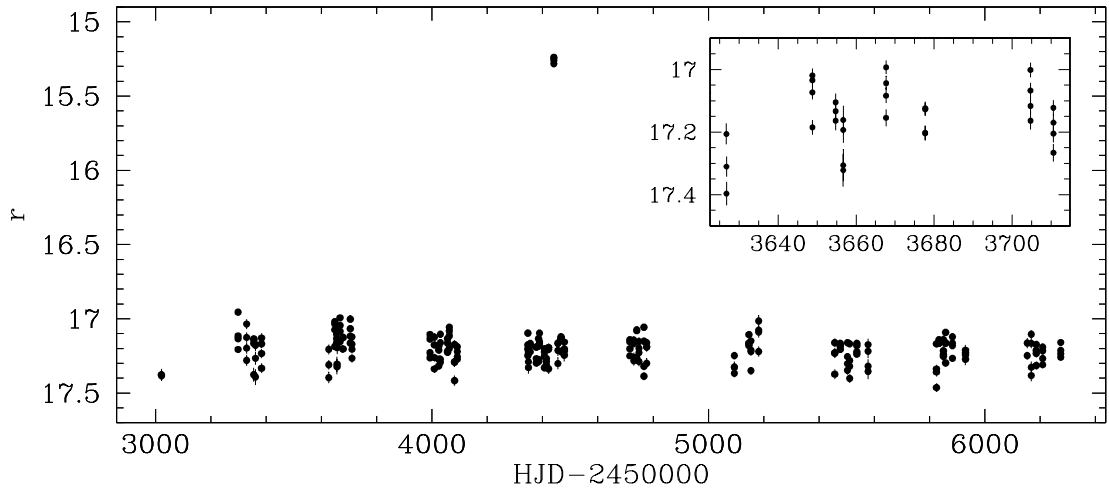}
\caption{CRTS light  curve of SDSS\,J0011-0647.  The  decrease from $r
  \sim 17$  to $r \sim  15$ is  due to SDSS\,J0011-0647  undergoing an
  outburst. A zoom-in  of 84 days is provided in  the top right corner
  of the figure. The data folded over the orbital period of the binary
  can be seen in Figure\,\ref{f-sol}.}
\label{f-catalina}
\end{figure}

Additional  photometry  of  SDSS\,J0011-0647 was  available  from  the
Catalina  Real-Time  Transient Survey  \citep[CRTS,][]{drakeetal09-1}.
We performed  differential photometry on the  reduced (bias-subtracted
and  flat-fielded)  CRTS images  following  the  routine described  by
\citet{parsonsetal13-1}.  Our  obtained $r$-band light curve  is shown
in  Figure\,\ref{f-catalina}.   The  CRTS   photometry  spans  a  time
interval  of  $\sim$9  years  and   reveals  an  increase  of  $\sim$2
magnitudes at  HJD $\sim$2454400  days.  We  interpret this  effect as
SDSS\,J0011-0647 undergoing an outburst, which clearly confirms the CV
nature of SDSS\,J0011-0647.

\section{Results}\label{results}

Here we make use of the spectroscopic and photometric data outlined in
the previous section  to obtain the orbital and  stellar parameters of
SDSS\,J0011-0647.

\begin{table}
\centering
\caption{\label{t-rvs}    \Lines{Na}{I}{8183.27,8194.81}\AA\,   radial
  velocities  of the  donor  star  in SDSS\,J0011-0647.   Heliocentric
  Julian dates, HJDs, are also provided.  The radial velocities folded
  over  the  orbital   period  are  shown  in  the   bottom  panel  of
  Figure\,\ref{f-sol}.}  \setlength{\tabcolsep}{1.6ex}
\begin{tabular}{cccccc}
\hline
\hline
 HJD            &  RV$_\mathrm{NaI}$  &    err &   HJD & RV$_\mathrm{H\alpha}$ &   err \\
  days       &  km/s           &        & days & km/s      & \\
\hline
 2455737.8675 &  -58.6 &  26.6 &  2456192.5668 &  337.7 &   9.3 \\
 2455737.8749 &  122.4 &  18.4 &  2456192.5815 &  239.1 &   9.3 \\
 2455737.8828 &  251.7 &  36.9 &  2456192.5888 &  123.0 &  10.8 \\
 2455737.8902 &  331.6 &  17.7 &  2456192.6040 & -173.4 &   7.7 \\
 2455737.8975 &  333.1 &  35.3 &  2456192.6257 & -244.1 &   7.8 \\
 2455737.9065 &  302.7 &  34.0 &  2456192.6332 & -171.9 &   7.7 \\
 2455737.9156 &  125.3 &  36.0 &  2456192.6405 &  -31.6 &  43.6 \\
 2455737.9264 &  -84.2 &  16.8 &  2456192.7190 & -269.5 &   8.2 \\
 2456192.5521 &  207.3 &  27.7 &  2456193.6488 &   75.1 &  13.9 \\
 2456192.5593 &  327.0 &  88.2 &  2456193.6892 &  163.8 &  11.6 \\
\hline
\end{tabular}
\end{table}

\subsection{Orbital period and semi-amplitude velocity of the donor star}

\begin{figure}[th]
\centering
\includegraphics[width=12cm]{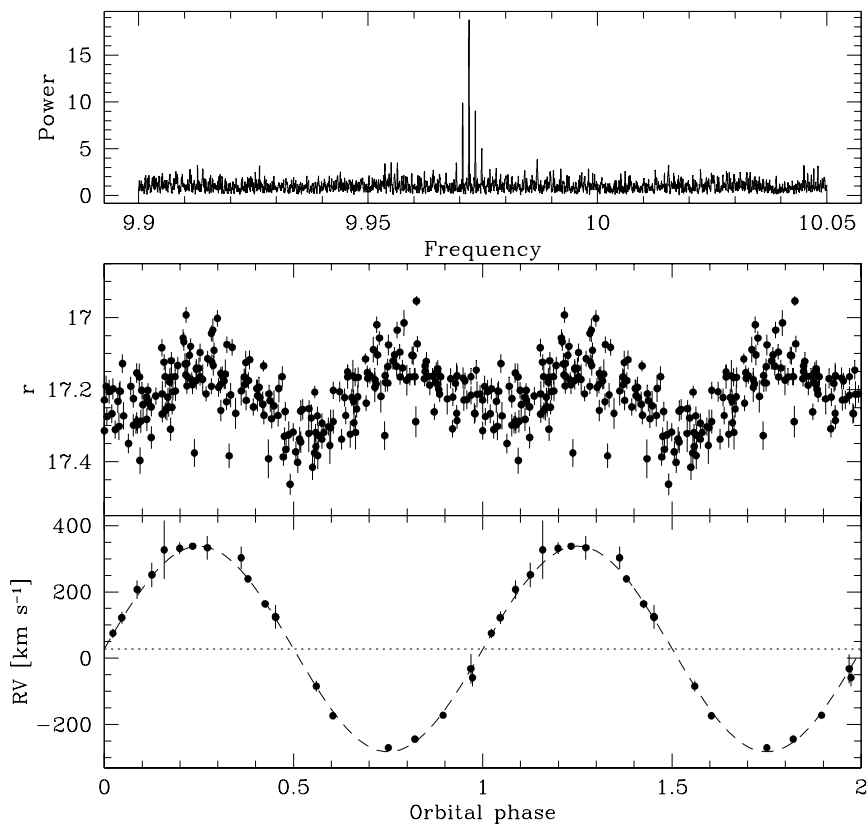}
\caption{Top:   power  spectra   obtained  applying   an  \textsf{ORT}
  periodogram to  the CRTS photometry of  SDSS\,J0011-0647, indicating
  an orbital  period of 2.40673  hours.  Middle: the CRTS  light curve
  folded       over      the       orbital      period.        Bottom:
  \Lines{Na}{I}{8183.27,8194.81}\AA\, (black solid dots, originated on
  the donor star)  radial velocities folded over our  adopted value of
  the orbital period and best sine-fit  to the data (dashed line). The
  horizontal dotted line represents the systemic velocity.}
\label{f-sol}
\end{figure}

We made  use of the  long time coverage  of the CRTS  photometric data
($\sim$9  years, see  Section\,\ref{s-phot})  to  determine a  precise
value  of the  orbital period  of  SDSS\,J0011-0647.  We  did this  as
followed.     We    first     run    an    \textsf{ORT}    periodogram
\citep{schwarzenberg-czerny96-1} to the CRTS  data and folded the CRTS
magnitudes over the best value of the orbital period obtained from the
periodogram, i.e.   2.40673 hours  (we excluded  in this  exercise the
CRTS    magnitudes    associated    to    the    recorded    outburst;
Section\,\ref{s-phot}).  The periodogram and corresponding CRTS folded
light   curve  are   shown   on   the  top   and   middle  panels   of
Figure\,\ref{f-sol}, respectively.  A  double-humped modulation can be
seen  in the  folded  light  curve which  we  identify as  ellipsoidal
modulation  of  the  Roche-lobe  filling donor.   We  then  fitted  an
ephemeris to a  set of 23 ($\phi=0 \pm 0.05$)  timings, which resulted
in

\begin{equation}
\mathrm{HJD} (\phi=0) = 2453383.578(1) + \mathrm{E}\times0.10028081(8),
\label{e-ephem}
\end{equation}

\noindent
where $\phi = 0$ refers to  the inferior conjunction of the donor star
and  HJD is  the heliocentric  Julian date  (in days).   We adopt  the
orbital period of SDSS\,J0011-0647 as  the one obtained from the above
photometric ephemeris, i.e. 2.40673946 hours.

We  measured the  radial velocities  of the  donor star  by fitting  a
second order polynomial  plus a double-Gaussian line  profile of fixed
separation   to  the   \Lines{Na}{I}{8183.27,8194.81}\AA\,  absorption
doublet      sampled      by      the      MIKE      spectra      (see
\citealt{rebassa-mansergasetal08-1}  for  details).   The  donor  star
radial  velocity  values   obtained  in  this  way   are  provided  in
Table\,\ref{t-rvs}  and  the radial  velocity  curve  folded over  the
adopted value  of the orbital period  is shown in the  bottom panel of
Figure\,\ref{f-sol}.  We performed a sine-fit of the form

\begin{equation}
\label{e-fit}
V_\mathrm{r} = K_\mathrm{don}\,\sin\left[\frac{2\pi(t-T_0)}{P_\mathrm{orb}}\right] +\gamma,
\end{equation}

\noindent
to the  \Lines{Na}{I}{8183.27,8194.81}\AA\, absorption  doublet folded
radial  velocity   curve  (dashed   line  in   the  bottom   panel  of
Figure\,\ref{f-sol}) to obtain the  systemic velocity $\gamma$ and the
semi-amplitude  velocity of  the  donor  star $K_\mathrm{don}$  (where
$T_0$   and    $P_\mathrm{orb}$   are    the   values    provided   by
Equation\,\ref{e-ephem}).   These values,  together  with our  adopted
value   of   the  orbital   period   and   $T_0$,  are   reported   in
Table\,\ref{t-param}.

\begin{table}
\centering
\caption{\label{t-param}   Stellar    and   orbital    parameters   of
  SDSS\,J0011-0647. In order of appearance are the orbital period, the
  systemic velocity,  the semi-amplitude  velocity of the  donor star,
  the semi-amplitude velocity of the  white dwarf, the mass ratio, the
  time of  inferior conjunction  of the  donor star,  the mass  of the
  white  dwarf,   the  mass  of   the  donor  star  and   the  orbital
  inclination.}  \setlength{\tabcolsep}{0.8ex}
\begin{tabular}{c}
\hline
\hline
\Porb [h]    =    2.40673946 $\pm$ 0.00000001 \\
$\gamma$ [km/s] = 28 $\pm$ 3\\
$K_\mathrm{don}$ [km/s] =  310 $\pm$ 4 \\
$K_\mathrm{wd}$ [km/s] = 100 $\pm$ 26\\
$q$ = 0.32 $\pm$ 0.08  \\
$T_0$ [HJD] =  2453383.578 $\pm$ 0.001 \\
$M_\mathrm{wd}$ [\Msun]  $>0.65\Msun$ \\
  $0.208\Msun< M_\mathrm{don}$  [\Msun] $<0.447\Msun$ \\
 $47< i$ [$^\circ$] $<70$\\
\hline
\end{tabular}
\end{table}

\subsection{Mass ratio and semi-amplitude velocity of the white dwarf}
\label{s-rot}

The high resolution  spectra obtained with MIKE allowed  us to measure
the  rotational  broadening  $V_\mathrm{rot}$ of  the  secondary  star
(affected by  the inclination  factor).  The  quantity $V_\mathrm{rot}
\times \sin i$ is related to the mass ratio $q$ via

\begin{eqnarray}
V_\mathrm{rot} \times \sin i = K_\mathrm{don} (q+1) {R_\mathrm{don} \over a}
\label{e-vsini}
\end{eqnarray}

\noindent \citep{horneetal86-1},  where for a Roche-lobe  filling star
$R_\mathrm{don}/a$ is a function of $q$ \citep{eggleton83-1}.  We used
this equation  to calculate  the mass  ratio $q$  and the  white dwarf
semi-amplitude  velocity $K_\mathrm{wd}$,  which  is  simply given  by
$K_\mathrm{wd} = q \times K_\mathrm{don}$.

We measured  the rotational broadening following  the method described
in   \citet{marshetal94-1}.   This   method  works   via  an   optimal
subtraction  routine,  in which  a  constant  times a  normalised  and
broadened  template  spectrum  is   subtracted  from  the  normalised,
orbitally corrected  target spectrum.  $V_\mathrm{rot}$  is determined
by finding the broadening factor  which minimises the $\chi^2$ between
this residual spectrum and a smoothed version of itself.

\begin{figure}[th]
\centering
\includegraphics[width=9cm, angle=-90]{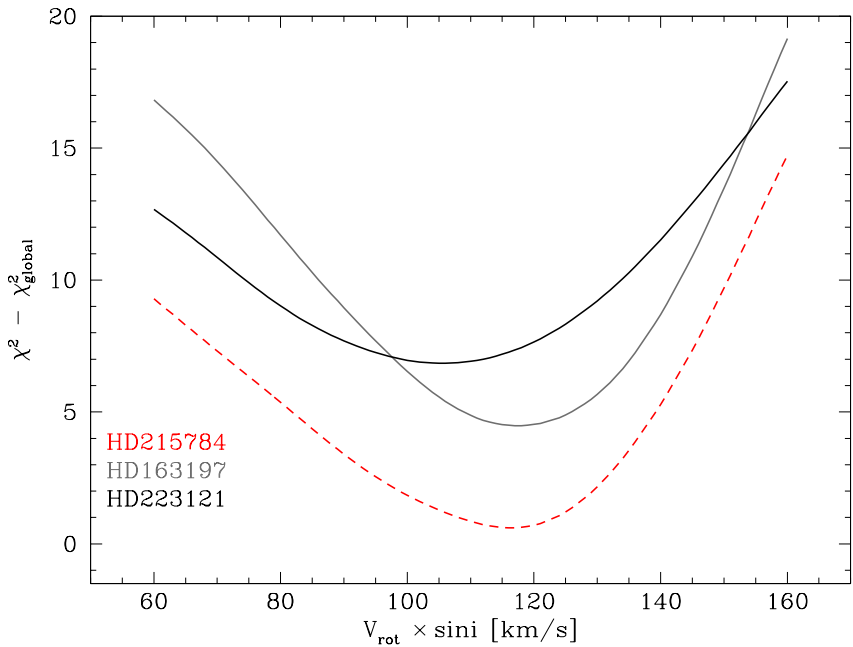}
\caption{Rotational broadening  measured fitting the  template spectra
  to the averaged observed spectrum  of SDSS\,J0011-0647 as a function
  of  $\chi^2-\chi^2_\mathrm{global}$, where  $\chi^2_\mathrm{global}$
  is  the global  $\chi^2$ minimum  ($\chi^2_\mathrm{min}$) among  all
  curves.  The best value of  the rotational broadening corresponds in
  each  case  to   $\chi^2_\mathrm{min}$  (see  Table\,\ref{t-templ}).
  Uncertainties  are assumed  as  the maximum  difference between  the
  values  below  $\chi^2-\chi^2_\mathrm{min}=1$.    The  best  fit  is
  obtained  when we  use HD215784  as  the template  star (red  dashed
  line).}
\label{f-vsini}
\end{figure}

Using   our   determinations    of   $K_\mathrm{don}$   and   $\gamma$
(Table\,\ref{t-param}), we applied an offset  to each MIKE spectrum to
remove the orbital  variation and then averaged the  spectra.  We also
obtained a  series of six  template stars  of known spectral  type and
luminosity class which were observed  as part of a different programme
(HD217880  (G8\,IV), HD215784  (K1\,IV),  HD163197 (K4\,IV),  HD220492
(G5\,V), HD223121  (K1\,V) and HD224287 (G8\,V);  D.  Steeghs, private
communication).    We   found   the   \Line{Fe}{I}{5455.61}\AA\,   and
\Line{Fe}{I}{5615.64}\AA\, absorption  lines to  be prominent  in both
the target  and all the template  spectra, and so use  these lines for
our  $V_\mathrm{rot}$ determination.   The  template  spectra cover  a
different wavelength range to the target, hence it was not possible to
use      more     obvious      spectral     features      like     the
\Lines{Na}{I}{8183.27,8194.81}\AA\,    absorption   doublet.     These
template spectra were  re-binned to match the target  spectrum, and we
broadened the  templates to account  for smearing  as a result  of the
orbital motion of the target.  For  each template star we then applied
a range of  rotational broadenings and computed  $\chi^2$ curves using
the optimal subtraction method.   Using the two \Ion{Fe}{I} absorption
lines we  obtained a  consistent measurement of  $V_\mathrm{rot}$ with
all six templates, but for the G stars the templates did not match the
target spectrum as well as the  K stars and subsequently the fits were
poorer,  thus we  excluded  them from  further consideration.   Figure
\ref{f-vsini} shows the resulting $\chi^2$ curves when considering the
\Line{Fe}{I}{5615.64}\AA\, line for our  K star templates.  The minima
of these  curves give the  preferred values of  $V_\mathrm{rot} \times
\sin i$ for  the cases where the donor  star in the CV is  of the same
type as the template star.

\begin{table}
\centering
\caption{\label{t-templ}  Template  spectra  used  for  measuring  the
  rotational  broadening of  the  donor star  in SDSSJ0011-0647.   The
  second column  indicates the  spectral type  of each  template.  The
  third,  fourth  and  fifth  columns  give  the  measured  rotational
  broadening and  the inferred $q$ and  $K_\mathrm{wd}$, respectively.
  In the  last column we  provide the quantity  $\chi_\mathrm{min}^2 -
  \chi_\mathrm{global}^2$, where $\chi_\mathrm{min}^2$  is the minimum
  of  the   $\chi^2$  curve   obtained  from  each   template  (Figure
  \ref{f-vsini})  and $\chi_\mathrm{global}^2$  is the  global minimum
  among all  $\chi_\mathrm{min}^2$.  HD\,215784 provides the  best fit
  ($\chi_\mathrm{min}^2   -    \chi_\mathrm{global}^2$=0,   see   also
  Figure\,\ref{f-vsinifit}).}  \setlength{\tabcolsep}{1.5ex}
\begin{tabular}{cccccc}
\hline
\hline
 Template  &  Sp  & $V_\mathrm{rot} \times \sin i$ & $q$ & $K_\mathrm{wd}$ & $\chi_{min}^2 -  \chi_{global}^2$ \\
           &      & km/s         &     &  km/s       &  \\
\hline
HD\,215784 & K1\,IV  & 117 $\pm$ 15 & 0.32 $\pm$ 0.08 & 100 $\pm$ 26 & 0 \\
HD\,163197 & K4\,IV  & 118 $\pm$ 12 & 0.33 $\pm$ 0.07 & 101 $\pm$ 20 & 3.9 \\
HD\,223121 & K1\,V   & 106 $\pm$ 17 & 0.26 $\pm$ 0.09 &  82 $\pm$ 28 & 6.2 \\
\hline
\end{tabular}
\end{table}

\begin{figure}[th]
\centering
\includegraphics[width=14cm]{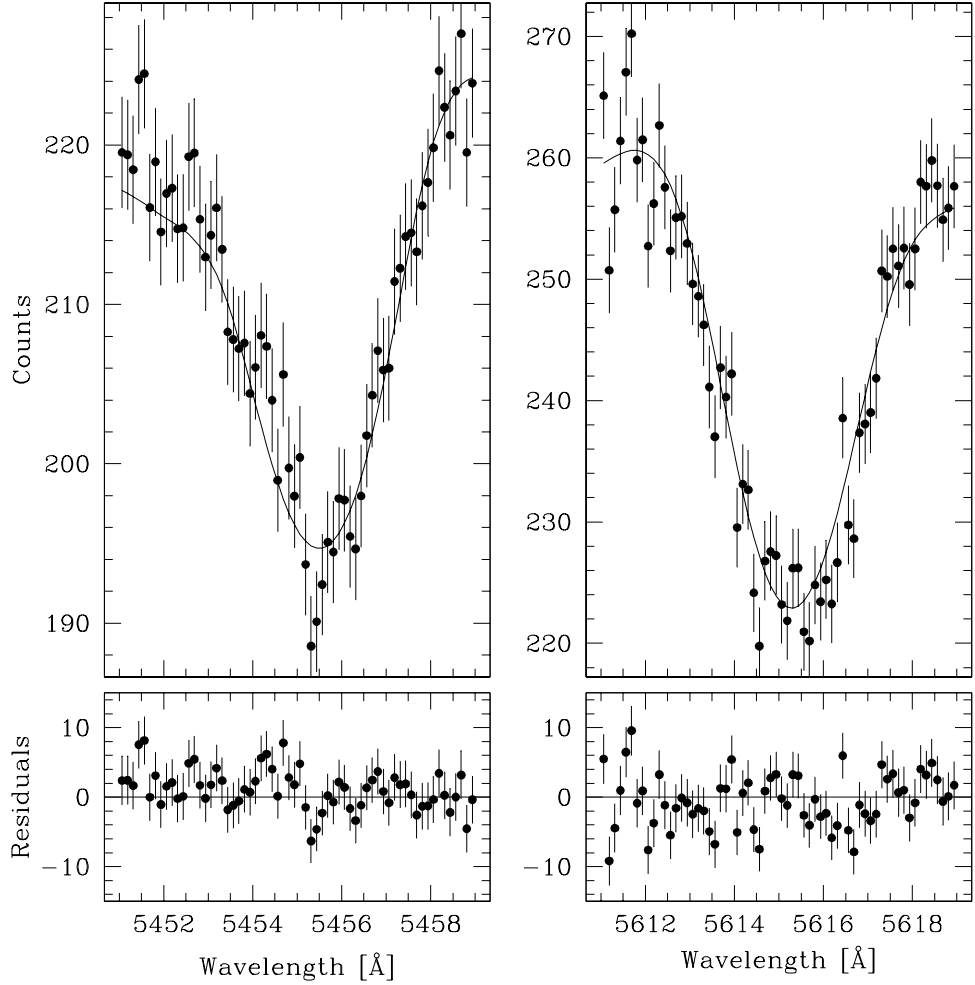}
\caption{Two \Ion{Fe}{I} absorption lines in the orbitally shifted and
  averaged spectrum of SDSS\,J0011-0647. The solid line plotted over
  the data points is the spectrum of our best fitting template star
  (HD\,215784), which has been artificially broadened by a factor
  $V_\mathrm{rot} \times \sin i = 117$ km/s and fitted to our data via
  an optimal subtraction routine as described in Section 3.2.}
\label{f-vsinifit}
\end{figure}

$q$   and   $K_\mathrm{wd}$   obtained  from   each   template   using
Equation\,\ref{e-vsini}  are given  in Table\,\ref{t-templ}.   In this
table   we   also   provide  the   quantity   $\chi_\mathrm{min}^2   -
\chi_\mathrm{global}^2$,   where   $\chi_\mathrm{min}^2$   gives   the
preferred value  of $V_\mathrm{rot} \times  \sin i$ for  each template
and  $\chi_\mathrm{global}^2$  is  the  global  $\chi_\mathrm{min}^2$.
Inspection  of Table\,\ref{t-templ}  and Figure  \ref{f-vsini} reveals
that  the  inferred $q$  and  $K_\mathrm{wd}$  values from  the  three
templates are in good agreement, however the best fit is obtained when
we  use HD\,215784  (see also  Figure\,\ref{f-vsinifit}), which  has a
spectral type of K1\,IV, as the template star.  We therefore adopt $q$
and $K_\mathrm{wd}$  as those obtained  using this star.   The adopted
$q$    and   $K_\mathrm{wd}$    values    are    also   reported    in
Table\,\ref{t-param}.

Based on  the multiplicative factor used  when we fit the  template to
the observed spectra we can also estimate the relative contribution of
the donor star, which is found to be  $37 \pm 10$ per cent when we use
HD\,215784  (spectral type  K1\,IV, our  best-fitting template  star).
Since  the contribution  from the  white dwarf  at this  wavelength is
expected to be small  ($\sim3-5$ per cent, see Section\,\ref{s-sptype}
for a further discussion on  this issue), the disk contribution should
be $\sim60-70$ per cent.  The  relative contribution of the donor star
is found to  be $30 \pm 9$ per  cent and $45 \pm 13$ per  cent when we
use our  two other template  stars HD\,163197 (K4\,IV)  and HD\,223121
(K1\,V), respectively, i.e. disk contributions of $\sim65-75$ per cent
and $\sim50-60$ per cent.

\subsection{Orbital inclination and stellar masses}
\label{s-masses}

With knowledge of the  orbital period, radial velocity semi-amplitudes
and mass  ratio (Table\,\ref{t-param}) we  can place a lower  limit on
the binary inclination using Kepler's laws and the fact that the white
dwarf should have a mass less than the Chandrasekhar mass,
\begin{eqnarray}
\Mwd = \frac{K_\mathrm{don} P_\mathrm{orb} (1+q)^2}{2 \pi G \sin{i}^3} < 1.4\Msun.
\end{eqnarray}
Using our  measured values  of $K_\mathrm{don}$,  $P_\mathrm{orb}$ and
$q$ gives  a limit on  the inclination of $i>47^\circ$.  This limit,
and the mass ratio,  also tells us that the mass of  the donor star is
$M_\mathrm{don}<0.447\Msun$.

\begin{figure}[th]
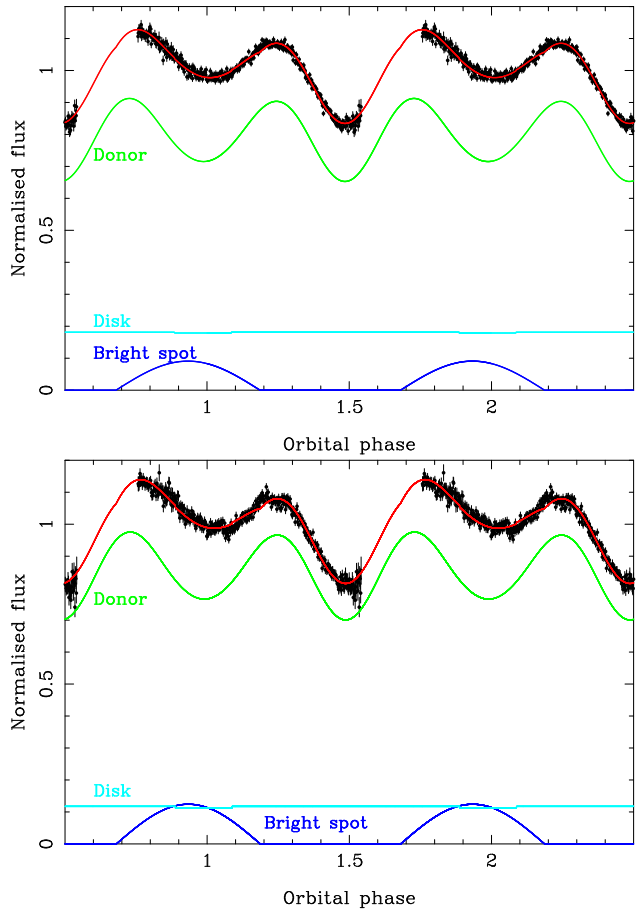

\centering
\includegraphics[width=6cm, angle=-90]{lcurve_fit_r.ps}
\includegraphics[width=6cm, angle=-90]{lcurve_fit_g.ps}
\caption{ULTRACAM $r'$ band (top) and  $g'$ band (bottom) light curves
  of SDSS\,J0011-0647.  Over-plotted are the best fit models (red). We
  also show the relative contributions of the donor star (green), disk
  (cyan) and bright spot (blue)  in these models.  These contributions
  vary  depending on  the considered  band.  Our  best models  have an
  inclination of  $67^\circ$ and the  disk is just eclipsed  (note the
  small dip in the disk light curves around phase 1).  At inclinations
  above $70^\circ$ this eclipse would  be deep enough to be detectable
  in our data.}
\label{f-lcfit}
\end{figure}

We can  place further constraints  on the system parameters  using our
Ultracam $g'$ and $r'$ band light curves (the $u'$ band light curve is
of low quality  and therefore not considered in  this exercise). These
light    curves,   folded    over    the    ephemeris   provided    in
Equation\,\ref{e-ephem},    are   shown    in   Figure\,\ref{f-lcfit}.
Unfortunately, we did not cover a  whole orbital cycle and hence there
is a small gap in phase coverage.  Inspection of Figure\,\ref{f-lcfit}
reveals the light  curves display a clear sinusoidal  variation with a
period of  half that  of the  binary itself.   This variation  is also
present in  the CRTS  light curve (Figure\,\ref{f-sol},  middle panel)
and is  caused by  the ellipsoidally  distorted donor  star presenting
different surface areas towards us throughout its orbit.  Furthermore,
it is clear that the minima at  phase 0.5 are much lower than at phase
1.

We  fitted   the  Ultracam  light   curves  using  a   code  developed
specifically   for   modeling   binaries   containing   white   dwarfs
\citep{copperwheatetal10-1}.  The  code accounts for  Roche distortion
and  we  included  both  a  disk and  bright  spot  component  in  our
model. The  disk contribution  is essentially constant  throughout the
orbit, whilst the  bright spot contribution rises around  phase 1 when
it is  directed towards the  observer and disappears around  phase 0.5
when its  light is moving  away from  us. Therefore, it  can naturally
account for  the differences in  the depths of  the two minima  in the
light curve,  by decreasing  the depth  of the  phase 1  minimum.  The
white dwarf in SDSS\,J0011-0647 contributes  a small amount of flux in
the $g'$  and $r'$  bands (see  Section\,\ref{s-sptype} for  a further
discussion on this issue).

We  set  the  mass  ratio  to  our  spectroscopically  measured  value
(Table\,\ref{t-param}) and fixed the donor  star's radius to its Roche
lobe.  We allowed the disk contribution  to vary and the bright spot's
contribution and  orientation to  vary as well.   We fitted  the light
curves  using  Levenberg-Marquardt  minimisation.   We  increased  the
inclination  in  steps  of  $1^\circ$  from  $47^\circ$  (the  minimum
inclination  based on  the white  dwarf's mass)  up to  $90^\circ$. We
emphasise that our aim is not  to determine the exact inclination, but
rather to set an upper limit  on the inclination, above which the disk
and bright spot are eclipsed (which is ruled out by the light curves).

The best fits to the  light curves are shown in Figure\,\ref{f-lcfit}.
We also  show the individual  components (donor star, disk  and bright
spot). We find that at inclinations larger than $70^\circ$ the eclipse
of the  disk and bright  spot is large enough  so that we  should have
detected it in  our light curves, hence we can  place this upper limit
on  the inclination.   Generally, at  inclinations where  there is  no
eclipse, the  light curve  is fairly  insensitive to  the inclination,
since  the contributions  from the  donor and  disk can  be varied  to
account   for  any   differences.   This   effect  can   be  seen   in
Figure\,\ref{f-lcfit},  where the  contributions  from the  individual
components resulting  from our best fits  to the two light  curves are
different.  Moreover,  we find  that even  for very  high inclinations
(i.e.   eclipsing)   the  disk   contributions  from  our   best  fits
($\sim10-20$ per  cent, depending on  the considered light  curve, see
Figure\,\ref{f-lcfit}) are lower  than the spectroscopically estimated
value ($\sim60-70$ per cent,  section\,\ref{s-rot}).  This fact, along
with  the possibility  of star-spots  on the  donor (which  could also
account for the varying depths of the light curve minima, and were not
included in  our model) means  that we are  unable to put  any tighter
constraints on the inclination.

Figure\,\ref{f-mfunc} shows  the masses  of both  the white  dwarf and
donor star as a function of inclination,  as well as our limits on the
inclination.   We   find  that  the   mass  of  the  white   dwarf  is
$\Mwd>0.65\Msun$,   and    the   mass    of   the   donor    star   is
$0.208\Msun<M_\mathrm{don}<0.447\Msun$. We  include these  values also
in Table\,\ref{t-param}.

\subsection{Spectral type of the secondary star}
\label{s-sptype}

The mass  of the donor  star of SDSS\,J0011-0647  is 0.21--0.45$\Msun$
(Section\,\ref{s-masses}),  which corresponds  to a  spectral type  of
M5.5--1.5  if we  assume a  zero-age main  sequence star  \citep[][see
  their  Table\,5]{rebassa-mansergasetal07-1}.  However,  our analysis
indicates  a   substantially  hotter  spectral  type   for  this  star
(Section\,\ref{s-rot}).   To investigate  this  hypothesis further  we
considered    a   set    of   nine    G,K,M   donor    star   template
spectra\footnote{Our main sequence template  spectra include six SEGUE
  \citep[the   SDSS   Extension   for   Galactic   Understanding   and
    Exploration,][]{yannyetal09-1}  G and  K stars  of spectral  types
  determined by the MILES  library \citep{sanchezetal06-1} and three M
  stars  from  the   library  of  \citet{rebassa-mansergasetal07-1}.},
combined them  with white dwarf model  spectra \citep{koester10-1} and
the  emission from  an isobaric  and  isothermal hydrogen  slab as  an
approximation   of    the   emission    from   the    accretion   disk
\citep{gaensickeetal99-1}, and compared  the resulting combined (donor
star plus accretion disk plus white dwarf) spectra to the SDSS optical
spectrum         and         Galaxy         Evolution         Explorer
\citep[$GALEX$,][]{martinetal05-1,    morrisseyetal05-1}   ultraviolet
fluxes of SDSS\,J0011-0647.

For  a given  donor  star template,  the  resulting combined  spectrum
depended on the white dwarf  effective temperature and surface gravity
(or mass) assumed, on the choice  of the accretion disk's gas pressure
and temperature and its optical depth  (where the latter relied on the
height of the isothermal hydrogen slab and the inclination of the disk
considered), and on the corresponding  contributions to the total flux
of the  three individual components  (white dwarf, accretion  disk and
donor star).  It is important to keep in mind that our aim here is not
to determine  all these physical characteristics  of SDSS\,J0011-0647,
as the  large number  of free parameters  would result  in significant
degeneracies, but rather to investigate whether or not combinations of
these parameters exist that match  well the observed spectrum, and use
these results to constrain the spectral type of the donor star.  Thus,
in order  to reduce the large  number of free parameters  involved, we
fixed the  white dwarf mass to  0.7\Msun\, and the inclination  of the
accretion disk to  67$^\circ$ (values that result  from the best-model
fit  to  our  Ultracam   light  curves,  Section\,\ref{s-masses})  and
considered $10^8$\,cm as the hydrogen  slab's height in all fits, thus
fixing also  the optical depth  of the accretion disk.   To facilitate
the  comparison the  observed and  combined  spectra (as  well as  the
$GALEX$ ultraviolet  fluxes) were normalised to  5615.64\AA, where the
contribution of the accretion disk  is estimated to be $\sim60-70$ per
cent (Section\,\ref{s-rot}).  For each donor star template we visually
explored a large number of composite models, spanning a representative
range in all free parameters, and computed a reduced $\chi^2$ for each
model.   In each  individual  fit the  $GALEX$  ultraviolet fluxes  of
SDSS\,J0011-0647  were  used to  directly  constrain  the white  dwarf
effective temperature and its contribution to the total combined flux.
We report  in Table\,\ref{t-specfit}  the parameters resulting  in the
lowest $\chi^2$  for each of  the donor  star templates, and  show the
corresponding  composite   models  along  with  the   observations  of
SDSS\,J0011-0647 in Figure\,\ref{f-stype}.

\begin{table}
\centering
\caption{\label{t-specfit}  Set  of  parameters that  allow  the  best
  combined fit (donor star template plus accretion disk model spectrum
  plus white  dwarf model spectrum; see  Figure\,\ref{f-stype}) to the
  optical spectrum and $GALEX$ ultraviolet fluxes of SDSS\,J0011-0647.
  In all  cases the  white dwarf  mass is fixed  to 0.7\Msun,  and the
  inclination of  the disk  to 67$^\circ$.   The contribution  of each
  component to the  total combined flux model at  5615.64\AA\, is also
  indicated.  The last  two columns indicate the  $\chi^2$ values that
  result from comparing  the best-fit combined model  and the observed
  spectrum    for     the    entire    optical     wavelength    range
  ($\chi_\mathrm{all}^2$)  and  for  the  blue range  of  the  optical
  spectrum    ($\chi_\mathrm{b}^2$,   3850-6000\AA),    respectively.}
\setlength{\tabcolsep}{1ex}
\begin{tabular}{ccccccccc}
\hline
\hline
   Template donor  & T$_\mathrm{WD}$ & P$_\mathrm{disk}$ & T$_\mathrm{disk}$ & WD & Donor & Disk & $\chi_\mathrm{all}^2$ & $\chi_\mathrm{b}^2$ \\
                   &               &                  &                & Contr. & Contr. & Contr. & & \\
                   & [K]         & [dyn/cm$^2$]   &  [K]          &  [\%]       &   [\%]     & [\%] \\
\hline
K4\,V              & 14000       &  300           & 5000          & 5           &  55        & 40 & 26  & 39 \\
G8\,IV             & 16000       &  500           & 5200          & 3           &  44        & 53 & 37  & 60 \\
K1\,IV             & 16000       &  800           & 5200          & 3           &  43        & 54 & 54  & 65 \\
K1\,V              & 16000       &  200           & 4600          & 3           &  51        & 46 & 46  & 79 \\
G8\,V              & 17000       &  200           & 4600          & 3           &  28        & 69 & 89 & 109 \\
K7\,V              & 14000       &  800           & 4800          & 5           &  13        & 82 & 65 & 112 \\
M0\,V              & 15000       &  800           & 4800          & 4           &  12        & 84 & 61 & 106 \\
M5\,V              & 15000       &  800           & 4800          & 4           &  1         & 95 & 89 & 159 \\
M3\,V              & 15000       &  800           & 4800          & 4           &  3         & 93 & 96 & 172 \\
\hline
\end{tabular}
\end{table}

Inspection of Figure\,\ref{f-stype} and Table\,\ref{t-specfit} reveals
that   the  best   combined   fits  to   the   observed  spectrum   of
SDSS\,J0011-0647  are  obtained  when  we use  the  K4\,V  and  G8\,IV
template stars, even though reasonably good fits are also obtained for
the  K1\,IV  and  K1\,V  templates. This  corroborates,  overall,  our
previous findings, i.e.  that the donor star has a  spectral type that
is much earlier  than expected for its mass.  The  quality of the fits
deteriorates when adopting  the later spectral types  K7\,V and M0\,V,
and  the worst  fits  are  obtained for  the  M3\,V,  M5\,V and  G8\,V
templates.  It is  worth noting that the  accretion disk contributions
are found to be $\sim40-55$ per  cent for the best-matching donor star
templates (falling between the  previous spectroscopic estimated value
of  $\sim60-70$ per  cent, Section\,\ref{s-rot},  and the  photometric
obtained  value  of  $\sim10-20$ per  cent,  Section\,\ref{s-masses}),
however  the   disk  contributions   increase  considerably   for  the
worse-fitting  donor star  templates  K7\,V, G8\,V,  M0\,V, M3\,V  and
M5\,V (Table\,\ref{t-specfit}).  This effect  is simple to understand:
when adopting a  donor star template that does not  match the observed
spectrum of  SDSS\,J0011-0647, our  fit decreases the  contribution of
the donor star to the composite model, the contribution of the disk is
increased to  compensate, and the  disk temperature is  optimized such
that  the  quasi-blackbody  spectrum  of  the  disk  approximates  the
observed data.   The most noticeable  feature that the  early-to-mid K
star  templates reproduce  well is  the strong  \Ion{Mg}{I} absorption
complex near 5150\AA.  This feature is weak, or absent, in the earlier
and later  donor stars, as  well as in  any accretion disk  model.  We
investigate the  diagnostic strength of the  \Ion{Mg}{I} absorption in
more  detail  by  comparing  the   $\chi^2$  values  obtained  in  the
3850-6000\AA\, wavelength range  (designated as $\chi_\mathrm{b}^2$ in
Table\,\ref{t-specfit}),  where the  observed  spectrum displays  both
spectral  features and  continuum.   Clearly,  the best-combined  fits
which provide the lowest $\chi^2$-fit values (those in which the donor
star template is  a K4\,V, G8\,IV, K1\,IV and K1\,V)  provide also the
lowest $\chi_\mathrm{b}^2$-fit values.

We  conclude that  spectral types  later than  K7\,V and  earlier than
G8\,V can  be ruled out for  the donor star of  SDSS\,J0011-0647. This
confirms in a robust way our  previous finding, i.e. that this star is
clearly too hot  for its mass. In the following  section we discuss in
detail  the  implications of  this  peculiarity  on the  evolution  of
SDSS\,J0011-0647.  We  also find  that, as  expected, the  white dwarf
contribution to the optical flux of SDSS\,J0011-0647 is small (3-5 per
cent at  5615.64\AA, Table\,\ref{t-specfit}),  and that  its effective
temperature is $\sim$14000-16000\,K.

\section{Discussion and conclusions}

SDSS\,J0011-0647  is   peculiar  CV  in   which  the  donor   star  is
substantially  hotter  (the  observations  are  best-matched  with  an
early-to-mid K  star) than expected for  its mass (0.21--0.45$\Msun$),
one of the most extreme cases  among CVs of its kind.  The present-day
configuration  of these  peculiar CVs  can be  explained if  the donor
stars underwent substantial nuclear  evolution before reaching contact
\citep{beuermannetal98-1,      baraffe+kolb00-1,     schenkeretal02-1,
  podsiadlowskietal03-1}.  This  is only possible  if at the  onset of
mass transfer the secondary  star of SDSS\,J0011-0647 was considerably
more  massive than  the mass  of canonical  CV donors.   Hence, it  is
almost  certain that  SDSS\,J0011-0647  passed  through a  substantial
phase of thermal-timescale  mass transfer, during which  it might have
been    expected     to    appear     as    a     super-soft    source
\citep{schenkeretal02-1}.   Super-soft  X-ray  sources  are  a  viable
channel        to       produce        type       Ia        supernovae
\citep[SN\,Ia,][]{vandenheuveletal92-1,             rappaportetal94-2,
  li+vandenheuvel97-1,  starrfieldetal04-1, distefano10-1}.   Although
SDSS\,J0011-0647 obviously  failed at  producing a SN\,Ia  during this
intense phase of  thermal transfer mass transfer,  its prior evolution
may help to  test models of SN\,Ia progenitors, i.e.   if we were able
to somehow  estimate the original  mass of  the white dwarf,  then the
present-day component masses would  constrain the accretion efficiency
of    the     white    dwarf    during    that     super-soft    phase
\citep{rodriguez-giletal09-1}.  However, the arguments are only likely
to be indirect.

Depending on the  degree of hydrogen depletion in  the center, evolved
donor   stars   are   not   expected  to   become   fully   convective
\citep{baraffe+kolb00-1}. Therefore they should  not be subject to the
same disruption  of magnetic braking  as the donor stars  in canonical
CVs.   These peculiar  CVs  are thus  not expected  to  detach in  the
orbital  period  gap  between  2-3   hours.   The  orbital  period  of
SDSS\,J0011-0647 is in the middle of the period gap, thereby providing
direct  observational evidence  supporting  the models  for these  hot
donor stars.   Arguably, this also  provides indirect support  for the
belief that  the period gap arises  as a consequence of  the secondary
stars in  normal CVs becoming  fully convective.  Future tests  of the
angular momentum loss in SDSS\,J0011-0647 may be able to test magnetic
braking in a previously unexplored regime.

\begin{figure}[th]
\centering
\includegraphics[width=9cm, angle=-90]{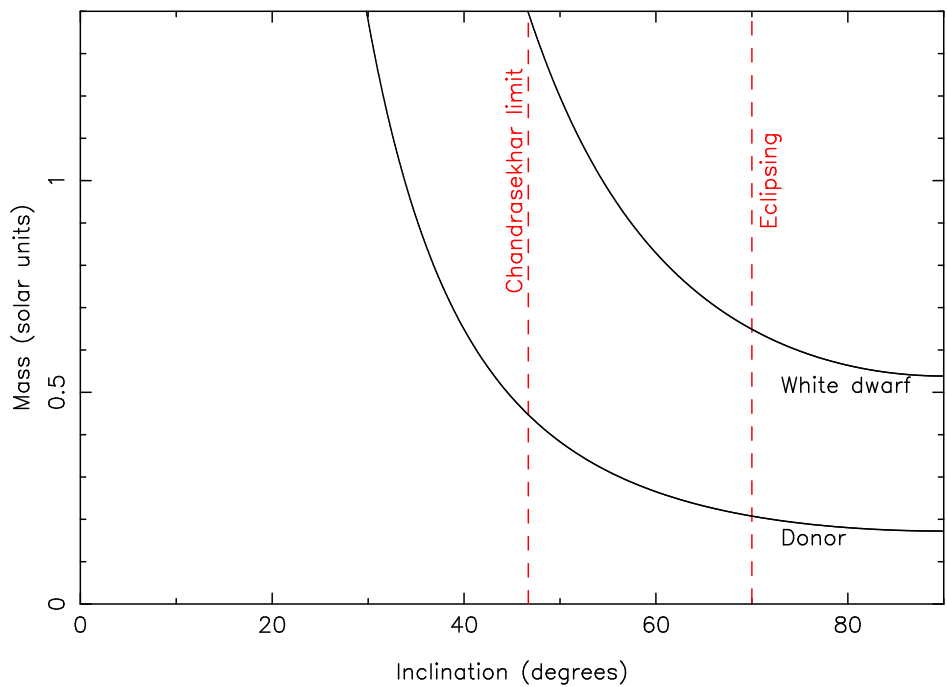}
\caption{Mass function plot for both the white dwarf and donor star in
  SDSS\,J0011-0647. A  lower limit on the  inclination of $47^\circ$
  is placed by  insisting that the white dwarf mass  be lower than the
  Chandrasekhar limit. The lack of any  disk or bright spot eclipse in
  the  light  curve  places  an  upper limit  on  the  inclination  of
  $70^\circ$.}
\label{f-mfunc}
\end{figure}

\begin{figure}[th]
\centering
\includegraphics[width=15cm]{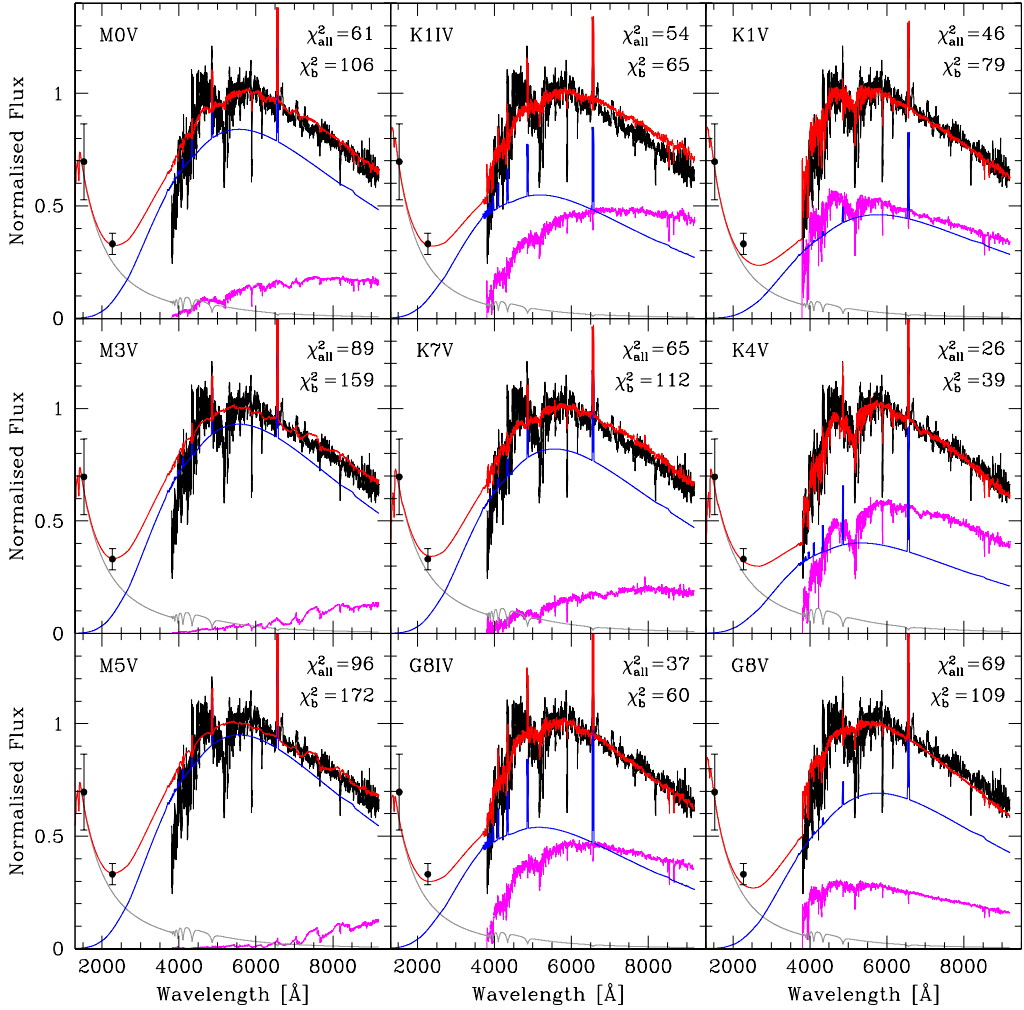}
\caption{Comparison of  the normalised  SDSS optical  spectrum (black)
  plus   $GALEX$   ultraviolet   fluxes    (black   solid   dots)   of
  SDSS\,J0011-0647 to a set of combined donor star template (magenta),
  white dwarf  model spectra (gray)  and accretion disk  model spectra
  (blue;  red for  the total  combined model,  also normalised).   The
  spectral types  of the  donor star templates  used and  the $\chi^2$
  that  result  from comparing  the  combined  model to  the  observed
  spectra  are  also  indicated (where  $\chi_\mathrm{all}^2$  is  the
  $\chi^2$   obtained    using   the   entire   optical    range   and
  $\chi_\mathrm{b}^2$ is  the $\chi^2$  obtained considering  only the
  3850-6000\AA\,  range).  A  spectral type  later than  K7\,V can  be
  ruled out  for the donor star,  which confirms in a  robust way this
  star is too hot for its inferred mass (Figure\,\ref{f-mfunc}).}
\label{f-stype}
\end{figure}

These CVs with evolved donor stars  are also predicted to later become
AM CVn-like  binaries \citep{schenkeretal02-1, podsiadlowskietal03-1}.
Further,   more   detailed,   measurements  of   the   properties   of
SDSS\,J0011-0647 would  thus help to refine  current models describing
this evolutionary phase. For example, future photometric near-infrared
observations  are   likely  to  constrain  considerably   the  orbital
inclination (and therefore  the stellar masses, Figure\,\ref{f-mfunc})
of  SDSS\,J0011-0647,  as the  disk  contribution  to the  ellipsoidal
modulation  is  essentially  negligible   at  these  wavelengths.   In
addition, it is  possible to indirectly measure the mass  of the white
dwarf from ultraviolet (e.g.  HST) spectroscopic observations.

Furthermore,  SDSS\,J0011-0647  may  help in  constraining  population
models   that   describe  the   evolution   of   these  peculiar   CVs
\citep{baraffe+kolb00-1, schenkeretal02-1,  podsiadlowskietal03-1}. It
is worth  noting that these  models have  not predicted many,  if any,
systems  with  similar  donor  temperatures  (or  spectral  types)  to
SDSS\,J0011-0647 at the  same orbital period.  This  may indicate that
the specific properties of SDSS\,J0011-0647 require unexpected or rare
initial   conditions.   For   example,  the   binary  grids   used  by
\citet{podsiadlowskietal03-1} may  not have  been able to  predict the
existence of  systems resembling  SDSS\,J0011-0647 either  because the
upper limit to the donor mass ($1.4 M_{\odot}$) was too low or because
the spacing in initial period  was insufficiently tight, i.e.\, either
the initial parameter  space was not large enough  or not sufficiently
finely-sampled.  In particular, since  the necessary orbital period at
contact may  have been  close to the  bifurcation period  beyond which
systems   would  become   wider   rather  than   tighter  (see,   e.g.
\citealt{pylyser+savonije88-1}),  some  careful   fine-tuning  may  be
required when setting up grids  of binary evolution sequences in order
to cover the part of  parameter space which produced SDSS\,J0011-0647.

Combining      SDSSJ170213.26+323954.1     \citep{littlefairetal06-1},
CSS\,J134052.0+151341 \citep{thorstensen13-1}  and the system  we have
discovered makes three  peculiar CVs containing hot  donor stars which
have near-identical  orbital periods  ($\approx 2.4$ hours).   To this
sample   we  could   add   the  recurrent   nova  IM\,Nor   (2.462\,h;
\citealt{woudt+warner03-2}),  as formation  models  for the  recurrent
novae with short orbital periods  typically involve a similar phase of
thermal-timescale  mass  transfer  onto the  white  dwarf  \citep[see,
  e.g.,][]{podsiadlowskietal03-2,    sarnaetal06-1}   (although    the
spectral type  of the  donor of  IM\,Nor is  unknown). Only  one other
peculiar     CV    containing     a    hot     donor,    QZ\,Serpentis
\citep{thorstensenetal02-2}, is known  in the period gap  (and that is
at the very short-period edge of  the gap).  This implies $\sim$1/2 of
the total currently-observed sample of  these peculiar CVs are located
in the period gap, and 4/5 of  those (if we include also IM\,Nor) with
nearly the  same orbital period.   Even though  we are subject  to low
number  statistics,  this concentration  of  systems  at a  particular
orbital period within the period gap may deserve further attention.

\acknowledgments {\it{Acknowledgments:} This work  is dedicated to our
  friend Dr.  Gisela  Andrea Romero, may she rest in  peace.  We thank
  Andrew  Drake  for providing  us  the  Catalina Real-Time  Transient
  Survey  images  of   SDSS\,J001153.08-064739.2,  Danny  Steeghs  for
  providing us  G and K  star template spectra observed  with Magellan
  Clay/MIKE,  and  the  anonymous  referee for  his/her  comments  and
  suggestions.    ARM   acknowledges   financial  support   from   the
  Postdoctoral  Science Foundation  of China  (grant 2013M530470)  and
  from the  Research Fund  for International  Young Scientists  by the
  National Natural  Science Foundation  of China  (grant 11350110496).
  SGP  acknowledges financial  support from  FONDECYT in  the form  of
  grant  number   3140585.   SJ   thanks  Philipp   Podsiadlowski  for
  discussions  and   CAS  and   NSFC  (Grant  No.    11350110324)  for
  support. MRS thanks  for support from FONDECYT  (grant 1141269). TRM
  acknowledges financial  support from  STFC grant  ST/L000733/1.  The
  research  leading to  these results  has received  funding from  the
  European  Research  Council  under   the  European  Union's  Seventh
  Framework Programme (FP/2007-2013) /  ERC Grant Agreement n.  320964
  (WDTracer).}

\newpage 

\end{document}